\documentclass[10pt,english,conference]{IEEEtran}
\usepackage{amsmath,amssymb,amsthm}
\usepackage[ruled,linesnumbered]{algorithm2e} 
\usepackage{algorithmicx}
\usepackage{algpseudocode}
\usepackage[square,comma,numbers,sort&compress]{natbib}
\usepackage[utf8]{inputenc}
\usepackage{hyperref}
\usepackage{mathtools} 
\usepackage{lipsum}
\usepackage{url}
\usepackage{array}
\usepackage{graphicx}
\usepackage[table]{xcolor}
\usepackage{xcolor,colortbl}
\usepackage{caption}
\usepackage{subcaption}
\usepackage{graphicx}
\usepackage{setspace}
\usepackage{comment}
\usepackage{amsfonts}
\usepackage{tikz}
\usetikzlibrary{arrows.meta, positioning}
\usepackage{enumerate}
\usepackage{mathtools}
\usepackage{siunitx}
\usepackage{tabularx}
\usepackage{todonotes}
\usepackage{setspace}
\usepackage{parskip}
\usepackage{fancybox}
\title{On the differential of the exponential map\thanks{This work is supported by a summer research scholarship from the Department of Electrical \& Computer Engineering, University of Maryland.}}
\author{
\IEEEauthorblockN{Debanjan Mallik}
\IEEEauthorblockA{\textit{Electrical and Computer Engineering} \\
\textit{University of Maryland, College Park}\\
College Park, MD, USA \\
dmallik@umd.edu}\
August 11, 2025}

\IEEEoverridecommandlockouts

\begin{document}

\maketitle
\date{August 10, 2025}

\begin{abstract}
\label{abstract}
   We study the time derivative of the matrix exponential $B(t)=\mathrm{exp}(A(t))$, where $A(t)$ is a time-parametrized curve in $\mathrm{Mat(n)}$. Starting from the Taylor series expansion, we derive a nested summation formula, which is then reformulated into a double summation. This expression is converted into an integral representation using Euler’s beta function, and further expressed in terms of Lie brackets and the adjoint representation. Later, we extend this to a family of problems and verify a well-known result. Finally, we establish connections to the Gateaux and Fréchet derivatives, proving the latter's existence. These results offer a unified and explicit framework for understanding the derivative of the exponential map.
\end{abstract}
\section{Introduction}
\label{intro}
In Lie Theory, the exponential map is a map from the Lie algebra  $\mathfrak{g}$ of a Lie group $G$ into $G$. If $G$ is a matrix Lie group, the exponential map reduces to the matrix exponential. This map is analytic in nature and it is formally denoted as $\exp: \mathfrak{g} \to G$. In this concise report, we will discuss the following problem on $\mathrm{Mat(n)}$, i.e. the space of all nxn matrices:
\begin{center}
\setlength{\fboxsep}{8pt}
\vspace{1ex}
\noindent
\begin{minipage}{0.85\columnwidth}
\centering
{Given a function $t \mapsto A(t)\in \mathrm{Mat}(n)$, $t\in [0,T]$; and, $B(t) = \exp(A(t))$, compute $\dot B(t)$, i.e., the time derivative of the matrix exponential.}
\end{minipage}
\vspace{1ex}
\end{center}
This problem leads to a useful result in Lie Theory, sometimes attributed to R.P. Feynman, but is likely a part of earlier work; see [4] and [5].

Section \ref{sum} begins with the Taylor expansion of $\exp$--- and, eventually, we find an initial form containing nested summations, and redefine the indices to reduce it to a double summation. In Section \ref{beta}, we introduce the Euler integrals of first and second kind, and utilize those ideas to derive the standard integral form. Section \ref{lie} expresses the earlier result using Lie brackets and little-$\mathrm{ad}$ notations, while Section \ref{family} extends the given problem to a family of problems. Section \ref{frechet} connects this problem to the notions of Gateaux and Fréchet derivatives.
\section{Nested and Double Summation Forms}
\label{sum}
Let, $A: [0, T] \mapsto \mathrm{Mat(n)}$ be continuously differentiable. We have, 
\begin{equation}
    B(t) = \exp(A(t)).
    \label{eq:1}
\end{equation}
Using Taylor series expansion, we can rewrite \eqref{eq:1} as,
\begin{equation*}
    B(t) = \sum_{k=0}^{\infty} \frac{1}{k!}(A(t))^k.
    \label{eq:2}
\end{equation*}
Now, the time derivative, $\dot B(t)$ is given by, 
  \begin{equation*}
    \dot{B}(t) = \frac{d}{dt} \sum_{k=0}^{\infty} \frac{1}{k!} \left( A(t) \right)^k.
    \label{eq:3}
\end{equation*} 
From the fundamental property of matrix exponentials, we know that $\exp$ converges, both absolutely and uniformly along with partial derivatives on any bounded set of matrices [3]. Thus, we are allowed to differentiate $\exp$ inside the sum,
\begin{equation}
    \dot B(t) = \sum_{k=0}^{\infty} \left(\frac{1}{k!}\frac{d}{dt}(A(t))^k\right).
    \label{eq:4}
\end{equation}
Since matrices do not commute in general, the chain rule for differentiation cannot be invoked here. Instead, we will treat the term $(A(t))^k$ as the product of $k$ identical factors, and apply the product rule,
\begin{equation*}
    \left(A(t)\right)^k = \underbrace{A(t)\, A(t)\, \dots \, A(t)}_{\text{\(k\ \text{factors}\)}}.
    \label{eq:5}
\end{equation*}
Differentiating each factor in turn while leaving the others unchanged, the $i^{th}$ term can be found,
\begin{equation*}
  term_i  =   \underbrace{A(t)\, \dots \, A(t)}_{\text{\(i-1\ \text{factors}\)}} \dot A(t) \underbrace{A(t)\, \dots \, A(t)}_{\text{\(k-i\ \text{factors}\)}}.
  \label{eq:6}
\end{equation*}
Setting $j = i - 1$(such that, $j$ runs from $0$ to $k-1$), we get the $i^{th}$ term as:
\begin{equation*}
    term_i = A(t)^j \dot A(t) A(t)^{k-1-j}.
    \label{eq:7}
\end{equation*}
Hence,
\begin{equation}
    \frac{d}{dt} (A(t))^k = \sum_{j=0}^{k-1} A(t)^j \dot A(t) A(t)^{k-1-j}.
    \label{eq:8}
\end{equation}
Using \eqref{eq:8}, we can rewrite \eqref{eq:4} as:
\begin{equation}
 \dot B(t) = \sum_{k=0}^\infty  \sum_{j=0}^{k-1}\frac{1}{k!} A(t)^j \dot A(t) A(t)^{k-1-j}.
 \label{eq:9}
\end{equation}
As discussed in Section \ref{intro}, we have expressed $\dot B(t)$ in a nested summation form in \eqref{eq:9}. Now, we aim to reduce it further into something more useful, such as a double summation. Let us redefine the index variables as:
\begin{equation*}
    p = j;\quad q = k-1-j.
    \label{eq:10}
\end{equation*}
Here, $p$ and $q$ independently run over $\{0,1, 2,\dots\}$. Moreover,
\begin{equation*}
    p+q = k-1,
    \label{eq:11}
\end{equation*}
or, 
\begin{equation*}
    k= p+q+1.
    \label{eq:12}
\end{equation*}
Substituting, we can rewrite the inner term in \eqref{eq:9} as:
\begin{equation*}
    \frac{1}{k!} A(t)^j \dot A(t) A(t)^{k-1-j} = \frac{1}{(p+q+1)!}A(t)^p \dot A(t) A(t)^q.
    \label{eq:13}
\end{equation*}
With \eqref{eq:9}, $\dot B(t)$ can finally be represented as a double summation,
\begin{equation}
    \dot B(t) = \sum_{p,q \geq 0}^{\infty} \frac{1}{(p+q+1)!}A(t)^p \dot A(t) A(t)^q.
    \label{eq:14}
\end{equation}
The factorial form in the double summation naturally leads to the beta integral, or more precisely, it relates to the beta-gamma relationship. We can get rid of the double summation, reducing it to exponential terms. We will discuss this further in section \ref{beta}. 

\section{Beta and Gamma Functions: Euler Integrals of First and Second Kind}
\label{beta}
Leonhard Euler introduced the ideas of gamma and beta functions in the first half of the $18\textsuperscript{th}$ century ($1720-30$s). The gamma function is also known as Euler integral of the second kind and is defined with a convergent improper integral,
\begin{equation*}
    \Gamma(x) = \int_{0}^\infty t^{x-1}e^{-t} dt;\quad x> 0.
    \label{eq:15}
\end{equation*}
The notation $\Gamma(.)$ is due to Legendre. \\
And, the beta function (Euler integral of the first kind) is defined as,
\begin{equation*}
    \mathcal{B}(x,y) =  \int_0 ^1 t^{x-1} (1-t)^{y-1} dt;\quad x,y>0.
    \label{eq:16}
\end{equation*}
A key result relating the beta integral to the gamma function is,
\begin{equation}
    \mathcal{B}(x,y) = \frac{\Gamma(x)\Gamma(y)}{\Gamma(x+y)}.
    \label{eq:17}
\end{equation}
Moreover, for any positive integer $n$,
\begin{equation}
    \Gamma(n) = (n-1)!.
    \label{eq:18}
\end{equation}
From \eqref{eq:17} and \eqref{eq:18}, taking $x= q+1$ and $y = p+1$, we can write,
\begin{equation}
    \int _0^1 (1-s)^p s^q ds = \frac{p!q!}{(p+q+1)!}.
    \label{eq:19}
\end{equation}
Since $(p+1)$ and $(q+1)$ are positive integers, $\Gamma(p+1) = p!, \Gamma(q+1) = q!, \text{and } \Gamma((p+1)+(q+1)) = (p+q+1)!$. Again, from \eqref{eq:19}, 
\begin{equation*}
    \frac{1}{(p+q+1)!} = \frac{1}{p!q!}\int_0^1 (1-s)^p s^q ds.
    \label{eq:20}
\end{equation*}
Suppressing the time argument,
\begin{align*}
    \frac{1}{(p+q+1)!} A^p \dot A A^q 
    &= \int_0^1 
        \frac{(1-s)^p}{p!} A^p \dot A 
        \frac{s^q}{q!} A^q \, ds.
\end{align*}
This is allowed since the order remains unchanged. Now, we can get rid of the double summation. Going back to the RHS in \eqref{eq:14}, 
\begin{multline*}
     \sum_{p,q \geq 0}^{\infty} \frac{1}{(p+q+1)!}A^p \dot A A^q = \\\int_0^1 \sum_{p=0}^\infty \frac{(1-s)^p}{p!} A^p \dot A \sum_{q=0}^\infty\frac{s^q}{q!} A^q ds.     
\end{multline*}
These are nothing but power series expansions for exponentials. Hence,
\begin{equation}
    \dot B(t) = \int_0^1 e^{(1-s)A(t)}\dot A(t) e^{sA(t)} ds.
    \label{eq:23}
\end{equation}
This is the final integral form.

\section{Lie Brackets/Adjoint Actions Form }
\label{lie}
In this section, we will derive $\dot B(t)$ in terms of adjoint actions. Little $\mathrm{ad}$ is defined as:
\begin{equation*}
    \mathrm{ad}_XY = XY -YX.
    \label{eq:24}
\end{equation*}
This is the adjoint action of a Lie algebra on itself. When thought of as an operation on matrices, $\mathrm{ad}_XY$ can also be written as a Lie bracket,
\begin{equation*}
    [X,Y] = XY- YX.
    \label{eq:25}
\end{equation*}
The Lie bracket operation is bilinear, antisymmetric and it satisfies the Jacobi identity,
\begin{equation*}
    [[X,Y], Z] + [[Z,X],Y] + [[Y,Z],X] = 0
    \label{eq:26}
\end{equation*}
Before delving deeper into the derivation, we will compute the powers of litte $\mathrm{ad}$,

\begin{equation*}
    \mathrm{ad}^0_XY = Y,
    \label{eq:27}
\end{equation*}
\begin{equation*}
    \mathrm{ad}^1_XY = \mathrm{ad}_XY = [X,Y],
    \label{eq:28}
\end{equation*}
\begin{equation*}
    \mathrm{ad}^2_XY = \mathrm{ad}_X(\mathrm{ad}_XY) = [X,[X,Y]] = [X,\mathrm{ad}_XY].
    \label{eq:29}
\end{equation*}
Generally, 
\begin{equation*}
    \mathrm{ad}^k_XY = [X, \mathrm{ad}^{k-1}_XY] = \mathrm{ad}_X(\mathrm{ad}^{k-1}_XY).
    \label{eq:30}
\end{equation*}
Now, we will derive an alternative representation of the final integral \eqref{eq:23}. Let, $u = (1-s)$; hence, $du = - ds$. Now,
\begin{equation*}
    \dot B(t) = - \int_{1}^0 e^{uA(t)} \dot A(t) e^{(1-u)A(t)}du,
    \label{eq:31}
\end{equation*}
or, 
\begin{equation*}
    \dot B(t) = \int_{0}^1 e^{uA(t)} \dot A(t) e^{(1-u)A(t)}du.
    \label{eq:32}
\end{equation*}
Furthermore, 
\begin{equation}
    \dot B(t) = \left(\int_{0}^1 e^{uA(t)} \dot A(t) e^{-uA(t)}du\right) e^{A(t)}.
    \label{eq:33}
\end{equation}
From \eqref{eq:23} and \eqref{eq:33}, we get,
\begin{multline*}
    \int_0^1 e^{(1-s)A(t)}\dot A(t) e^{sA(t)} ds = \\\left(\int_{0}^1 e^{sA(t)} \dot A(t) e^{-sA(t)}ds\right) e^{A(t)}.
    \label{eq:34}
\end{multline*}
Wilcox(1967)[2] states the well-known expansion,
\begin{equation}
    e^A Be^{-A} = B + [A,B] + \frac{1}{2!} [A[A,B]] + \dots 
    \label{eq:35}
\end{equation}
We notice that the RHS in \eqref{eq:35} is a series containing the powers of little-$\mathrm{ad}$. Hence, 
\begin{equation}
    e^A Be^{-A} = \sum_{k=0}^\infty \frac{1}{k!} \mathrm{ad}^k_AB. 
    \label{eq:36}
\end{equation}
Moreover, we recall,
\begin{equation}
    \mathrm{ad}^k_{sA}B = s^k\mathrm{ad}^k_AB,
    \label{eq:37}
\end{equation}
and,
\begin{equation}
    e^{sA} Be^{-sA} = \sum_{k=0}^\infty \frac{1}{k!} \mathrm{ad}^{k}_{sA}B.
    \label{eq:38}
\end{equation}
From \eqref{eq:33} - \eqref{eq:38}, we get,
\begin{equation}
    \dot B(t) = \left(\int_0^1 \sum_{k=0}^\infty \frac{1}{k!} \mathrm{ad}^{k}_{sA(t)}\dot A(t) ds\right) e^{A(t)}.
    \label{eq:39}
\end{equation}
From \eqref{eq:37}, we can write,
\begin{equation*}
    \dot B(t) = \left(\int_0^1 \sum_{k=0}^\infty \frac{s^k}{k!} \mathrm{ad}^k_{A(t)}\dot A(t) ds\right) e^{A(t)}.
    \label{eq:40}
\end{equation*}
Using the integral form derived in Section~\ref{beta}, we can also rewrite \eqref{eq:39} as:
\begin{equation*}
    \dot B(t) = e^{A(t)}\left(\int_{0}^1 \sum_{k=0} ^\infty \frac{1}{k!}\mathrm{ad}^k_{-sA(t)}\dot A(t) ds\right).
    \label{eq:41}
\end{equation*}
Again,
\begin{equation*}
    \dot B(t) = e^{A(t)}\left(\int_0^1 \sum_{k=0}^\infty \frac{(-1)^k}{k!} s^k \mathrm{ad}^k_{A(t)} \dot A(t) ds\right),
    \label{eq:42}
\end{equation*}
and, 
\begin{equation}
    e^{-\mathrm{ad}_X} = \sum_{k=0}^\infty \frac{(-1)^k}{k!} \mathrm{ad}^k_X.
    \label{eq:43}
\end{equation}
Hence,
\begin{equation*}
    \dot B(t) = e^{A(t)}\left(\int_0^1 e^{-s\cdot \mathrm{ad}_{A(t)}} \dot A(t) ds\right).
    \label{eq:44}
\end{equation*}
Now, we will derive a simpler result that will be beneficial in deriving the intended form of $\dot B(t)$. \\
From \eqref{eq:43}, we write,
\begin{equation*}
    \mathbb{I} - e^{-\mathrm{ad}_X} = \mathbb{I} - \left(\mathbb{I} + \sum_{k=1}^\infty \frac{(-1)^k}{k!} \mathrm{ad}^k_X\right).
    \label{eq:45}
\end{equation*}
Here, $\mathbb{I}$ denotes the identity matrix.\\
Or, 
\begin{equation}
    \mathbb{I} - e^{-\mathrm{ad}_X} = -\mathrm{ad}_X\left(\sum_{k=1}^\infty \frac{(-1)^k}{k!} \mathrm{ad}^{k-1}_X\right).
    \label{eq:46}
\end{equation}
Letting $\ell= k-1$, we rewrite \eqref{eq:46} as:
\begin{equation}
    \mathbb{I} - e^{-\mathrm{ad}_X} = \mathrm{ad}_X\left(\sum_{\ell=0}^\infty\frac{(-1)^\ell}{(\ell+1)!} \mathrm{ad}^\ell_X\right).
    \label{eq:47}
\end{equation}
From \eqref{eq:47}, we obtain formally,
\begin{equation}
    \frac{\mathbb{I} - e^{-\mathrm{ad}_X}}{\mathrm{ad}_X} = \sum_{\ell=0}^\infty\frac{(-1)^\ell}{(\ell+1)!} \mathrm{ad}^\ell_X.
    \label{eq:48}
\end{equation}
Now,
\begin{equation}
    \int_0^1 e^{-s\cdot\mathrm{ad}_{A(t)}}\dot A(t)ds = \int_0^1\left(\sum_{k=0}^\infty \frac{(-s)^k}{k!} \mathrm{ad}^k_{A(t)}\dot A(t)\right) ds.
    \label{eq:49}
\end{equation}
In case of uniform convergence, summation and integral can be interchanged. Therefore,
\begin{equation}
    \int_0^1 e^{-s\cdot\mathrm{ad}_{A(t)}}\dot{A}(t)ds = \sum_{k=0}^\infty \left(\int_0^1 \frac{(-s)^k}{k!} ds\right)\mathrm{ad}^k_{A(t)}\dot A(t).
    \label{eq:50}
\end{equation}
Integrating term-by-term, 
\begin{equation}
    \int_0^1 e^{-s\cdot \mathrm{ad}_{A(t)}}\dot A(t)ds = \sum_{k=0}^\infty \frac{(-1)^k}{(k+1)!} \mathrm{ad}^k_{A(t)}\dot A(t).
    \label{eq:51}
\end{equation}
From \eqref{eq:48}, we define formally,
\begin{equation}
    \dot B(t) = e^{A(t)} \frac{\mathbb{I} - e^{-\mathrm{ad}_{A(t)}}}{\mathrm{ad}_{A
    (t)}}\dot A(t).
    \label{eq:52}
\end{equation}
This is our intended form, expressed in terms of little -$\mathrm{ad}$ notation. This formula was first proved by Friedrich Schur($1891$) [4], and later by Henri Poincaré($1899$) [5].
\subsection{An Alternative Proof}
This result \eqref{eq:52} can also be proved using an alternative approach, as given in [1]. 
Under the hypothesis of $A\equiv A(t)$ being continuously differentiable in the given range, we set,
\begin{equation}
    Y(s,t) = e^{-sA(t)}\frac{\partial{}}{\partial{t}} e^{sA(t)}. 
    \label{eq:53}
\end{equation}
Differentiating with respect to s,
\begin{equation*}
    \frac{\partial{Y}}{\partial{s}} = e^{-sA}(-A)\frac{\partial{}}{\partial{t}} e^{sA}  + e^{-sA}\frac{\partial{}}{\partial{t}}(Ae^{sA}).
    \label{eq:54}
\end{equation*}
Mixed partial derivatives commute.\\
Or,
\begin{equation*}
  \frac{\partial{Y}}{\partial{s}} =  e^{-sA}(-A)\frac{\partial{}}{\partial{t}} e^{sA}  + e^{-sA} \dot A e^{sA} + e^{-sA}A\frac{\partial{}}{\partial{t}} e^{sA}.
  \label{eq:56}
\end{equation*}
The first term and the third term in RHS cancel out each other,
\begin{equation}
   \frac{\partial{Y}}{\partial{s}} = e^{-sA} \dot A e^{sA}.
   \label{eq:56}
\end{equation}
From \eqref{eq:38}, we can rewrite \eqref{eq:56} as:
\begin{equation*}
    \frac{\partial{Y}}{\partial{s}} = \sum_{k=0}^\infty\frac{1}{k!}\mathrm{ad}^k_{-sA}\dot A,
    \label{eq:57}
\end{equation*}
or, 
\begin{equation*}
    \frac{\partial{Y}}{\partial{s}} = e^{\mathrm{ad}_{-sA}} \dot A.
    \label{eq:58}
\end{equation*}
From fundamental theorem of calculus,
\begin{equation*}
    \int_{0}^1 \frac{\partial}{\partial s} Y(s,t) ds = Y(1,t)- Y(0,t).
    \label{eq:59}
\end{equation*}
From \eqref{eq:53},
\begin{equation*}
    e^{-A} \frac{d}{dt}{\left(e^A\right)} = Y(1,t).
    \label{eq:60}
\end{equation*}
As $Y(0,t) = 0$, we can write,
\begin{equation*}
      e^{-A} \frac{d}{dt}{\left(e^A\right)} = \int_{0}^1 \frac{\partial}{\partial s} Y(s,t) d s,
\label{eq:61}    
\end{equation*}
and,
\begin{equation*}
    \frac{\partial{Y}}{\partial{s}} = e^{-s\cdot\mathrm{ad}_{A}}\dot A.
    \label{eq:62}
\end{equation*}
From \eqref{eq:49} - \eqref{eq:51}, 
\begin{equation*}
    e^{-A} \frac{d}{dt}{\left(e^A\right)} = \sum_{k=0}^\infty\frac{(-1)^k}{(k+1)!} \mathrm{ad}^k_A \dot A.
    \label{eq:63}
\end{equation*}
Thus,
\begin{equation*}
    \dot B(t) = \frac{d}{dt}\left(e^{A(t)}\right) = e^A(t)  \left(\sum_{k=0}^\infty\frac{(-1)^k}{(k+1)!} \mathrm{ad}^k_{A(t)} \right)\dot A(t).
    \label{eq:64}
\end{equation*}
From \eqref{eq:48},
\begin{equation*}
    \dot B(t) = e^{A(t)} \frac{\mathbb{I} - e^{-\mathrm{ad}_{A(t)}}}{\mathrm{ad}_{A
    (t)}}\dot A(t) .
    \label{eq:65}
\end{equation*}
This is exactly the result we derived earlier in \eqref{eq:52}.

\section{Extension to a Family of Problems}
\label{family}
In this section, we will extend the integral form \eqref{eq:23} to a family of problems, and verify this extension as stated in [2]. \\
\textbf{Lemma:}\text{(Wilcox [2])} If the operator $H$ is a function of a parameter $\lambda$, $H \equiv H(\lambda)$, then,
\begin{equation}
    \frac{\partial}{\partial \lambda} e^{-\beta H} = -\int_0^\beta e^{-(\beta - u)H}  \frac{\partial H}{\partial \lambda} e^{-uH} du
    \label{eq:66}.
\end{equation}
The factor $-\beta$ makes the above equation a more general form of \eqref{eq:23}. In our report, time($t$) is the analogous parameter to $\lambda$. $A\equiv A(t)$ is the matrix-valued function, similar to $H\equiv H(\lambda)$. Now, let us verify \eqref{eq:66}. At first, we will show that LHS = RHS at $\beta = 0$.\\
\textbf{Verification: }At  $\beta = 0$, LHS:
\begin{equation*}
    \frac{\partial}{\partial \lambda} e^{-\beta H} \Big|_{\beta = 0} = \frac{\partial}{\partial \lambda} \mathbb{I} = 0,
    \label{eq:67}
\end{equation*}
and, RHS:
\begin{equation*}
    \int_0^\beta e^{-(\beta - u)H}  \frac{\partial H}{\partial \lambda} e^{-uH} du \Big|_{\beta = 0} = \int_0^\beta e^{ uH} \frac{\partial H}{\partial \lambda} e^{-uH} du = 0.
    \label{eq:68}
\end{equation*}
Hence, LHS = RHS at $\beta = 0$. Let, 
\begin{equation}
    F(\beta, \lambda) = \frac{\partial}{\partial \lambda} e^{-\beta H(\lambda)},
    \label{eq:69}
\end{equation}
and, 
\begin{equation}
    G(\beta, \lambda) =  -\int_0^\beta e^{-(\beta - u)H}  \frac{\partial H}{\partial \lambda} e^{-uH} du.
    \label{eq:70}
\end{equation}
We need to be careful about the negative sign in RHS. Differentiating \eqref{eq:69} with respect to $\beta$,
\begin{equation*}
    \frac{\partial}{\partial \beta} F(\beta, \lambda) = \frac{\partial}{\partial \beta} \frac{\partial}{\partial \lambda} e^{-\beta H(\lambda)} = \frac{\partial}{\partial \lambda} \frac{\partial}{\partial \beta} e^{-\beta H(\lambda)},
    \label{eq:71}
\end{equation*}
or,
\begin{align*}
     \frac{\partial}{\partial \beta} F(\beta, \lambda) = \frac{\partial}{\partial \lambda} \left\{e^{-\beta H(\lambda)}  (-H(\lambda))\right\} \\=  \frac{\partial}{\partial \lambda} \left\{(-H(\lambda))e^{-\beta H(\lambda)}  \right\},
    \label{eq:72_73}
\end{align*}
or,
\begin{equation*}
    \frac{\partial}{\partial \beta} F(\beta, \lambda) = (-H)(\partial_{\lambda}e^{-\beta H}) - (\partial_\lambda H)e^{-\beta H},
    \label{eq:74}
\end{equation*}
or, 
\begin{equation}
    \frac{\partial}{\partial \beta} F(\beta, \lambda) = (-H) F - (\partial_\lambda H)e^{-\beta H}.
\label{eq:75}
\end{equation}
Now, differentiating \eqref{eq:70} with respect to $\beta$,
\begin{equation*}
    \frac{\partial}{\partial \beta} G(\beta, \lambda) = -\frac{\partial}{\partial \beta}  \int_0^\beta e^{-(\beta - u)H}  \frac{\partial H}{\partial \lambda} e^{-uH} du.
\end{equation*}
Using the Leibniz rule, 
\begin{multline*}
  \frac{\partial}{\partial \beta} G(\beta, \lambda) = -e^ {-(\beta - \beta)H}(\partial_{\lambda}H) e^{-\beta H} \\-\int_0^\beta\frac{\partial}{\partial \beta}\{e^{-(\beta - u) H} (\partial_\lambda H) e^{-uH}\} du,
\end{multline*}
or, 
\begin{multline*}
      \frac{\partial}{\partial \beta} G(\beta, \lambda) = -(\partial_{\lambda}H) e^{-\beta H} \\- \int_0^\beta (-H) e^{-(\beta - u)H}(\partial_\lambda H) e^{-uH} du,
      \end{multline*}
or,
\begin{multline}
     \frac{\partial}{\partial \beta} G(\beta, \lambda) = -(\partial_{\lambda}H) e^{-\beta H} + (-H)G \\= (-H)G  -(\partial_{\lambda}H) e^{-\beta H}.
     \label{eq:79}
\end{multline}
From \eqref{eq:75} and \eqref{eq:79}, we have,
\begin{equation*}
   \frac{\partial}{\partial\beta}(F-G) =  -H(F-G) \text{ with }(F-G)\Big|_{\beta = 0} = 0.
   \label{eq:80}
\end{equation*}
The only possible solution with zero initial data is $F-G \equiv 0.$ Hence,
\begin{equation*}
    F(\beta, \lambda) = G(\beta, \lambda),
    \label{eq:81}
\end{equation*}
and, \eqref{eq:66} holds.

\section{Connection to Gateaux and Fréchet Derivatives}
\label{frechet}
Let us recall the concepts of Gateaux and Fréchet derivatives first. Let $X$ be a vector space and let $Y$ be a normed linear space equipped with the norm $||\cdot||$. Let, $f:X \to Y$ be an operator, not necessarily linear. If the limit,
\begin{equation*}
    \partial f(x; h) = \lim_{\varepsilon \to 0} \frac{1}{\varepsilon}\left(f\left(x + \varepsilon h\right) - f(x)\right)
    \label{eq:82}
\end{equation*}
exists, then $\partial f(x; h)$ is called the Gateaux differential of $f$ at $x$ with an increment $h$. If the limit exists $\forall h \in X$, $f$ is said to be Gateaux differentiable and, the operator $h \mapsto \partial f(x; h)$ is called the Gateaux derivative operator. We know,
\begin{equation*}
    \lim_{\varepsilon \to 0} \frac{1}{\varepsilon}\left(f(x + \varepsilon h) - f(x)\right) = \frac{d}{d\varepsilon} f\left(x+ \varepsilon h\right) \big|_{\varepsilon = 0}.
    \label{eq:83}
\end{equation*}

Similarly, let $X$ be normed and $U \subset X$ be an open subset containing the point $x$. If there exists a linear, bounded map $Df(x;\cdot) : X\to Y(\text{hence, } h\mapsto Df(x;h))$ such that,
\begin{equation*}
    \lim_{||h||\to0} \frac{||f(x+h) - f(x) -Df(x;h)||_{Y}}{||h||_X} = 0.
    \label{eq:84}
\end{equation*}
Then, $f$ is said to be Fréchet differentiable at $x\in U \subset X$ and $Df(x;h)$ is called the Fréchet differential. The operator $Df(x; \cdot)$ is the Fréchet derivative operator.\\
\textbf{Lemma:}(Equality of differentials [7]) If the Fréchet differential $Df(x;h)$ exists at $x$, with increment $h$, so does the Gateaux differential $\partial f(x;h)$, and they must be equal. 

Now, coming back to our problem, let us consider $X = Y= \mathrm{Mat(n)}$. We can treat the exponential map $\exp$ as $f$. Thus, $\exp: U \subset^{open} \mathrm{Mat(n)} \to \mathrm{Mat(n)}$. The Gateaux differential of $\exp$ at $A$, in the direction $H$ is,
\begin{equation}
    \partial \exp(A;H) = \lim_{\varepsilon\to0} \frac {e^{(A+\varepsilon H)} - e^{A}}{\varepsilon} = \frac{d}{d\varepsilon}e^{(A+\varepsilon H)} \big|_{\varepsilon = 0}
.    \label{eq:85}
\end{equation}
Note that this $H$ in \eqref{eq:85} is the increment in $\mathrm{Mat(n})$. It is different from $H \equiv H(\lambda)$ mentioned in the lemma in Section \ref{family}.
Now, for $A(\varepsilon) := A + \varepsilon H,$
\begin{equation*}
e^{A(\varepsilon)} = \sum_{k=0}^\infty \frac{1}{k!}\left(A+\varepsilon H\right)^k, 
\label{eq:86}
\end{equation*}
and, 
\begin{equation*}
    \frac{d}{d\varepsilon}e^{A(\varepsilon)} = \sum_{k=0}^\infty \frac{1}{k!}\frac{d}{d\varepsilon}(A+\varepsilon H)^k .
    \label{eq:87}
\end{equation*}
Again,
\begin{equation}
    \frac{d}{d\varepsilon}(A+\varepsilon H)^k = \sum_{j=0}^{k-1} (A+\varepsilon H)^j H(A+\varepsilon H)^{k-1-j}.
    \label{eq:88}
\end{equation}
At $\varepsilon =0$, \eqref{eq:88} becomes,
\begin{equation*}
    \frac{d}{d\varepsilon}(A+\varepsilon H)^k\big|_{\varepsilon = 0} = \sum_{j=0}^{k-1} A^j HA^{k-1-j}.
    \label{eq:89}
\end{equation*}
Following the approach in Sections {\ref{sum}} and \ref{beta}, we can conclude,
\begin{equation}
   \partial \exp(A;H)= \frac{d}{d\varepsilon}e^{A(\varepsilon)}\big|_{\varepsilon = 0} = \int_0^1 e^{(1-s)A}H e^{sA} ds.
    \label{eq:90}
\end{equation}
This is the Gateaux derivative, and as the above lemma states, this also is our Fréchet candidate. Now, let us find out if the Fréchet derivative exists. In other words, we need to show that the error term $e^{(A+ H)} - e^A - \partial \exp(A;H)$ is $o(||H||)$ as $||H||\to0$.
We know,
\begin{equation*}
    e^{(A+H)} - e^A =\sum_{k=0}^\infty\frac{1}{k!} (A+H)^k  - \sum_{k=0}^\infty \frac{1}{k!}A^k.
    \label{eq:91}
\end{equation*}
Now, let us compute $(A+H)^k - A^k$. Let, $Z= A+H$,
\begin{equation*}
    Z^k - A^k = Z\cdot Z^{k-1} - A\cdot A^{k-1},
    \label{eq:92}
\end{equation*}
or,
    \begin{equation}
      Z^k - A^k  = (Z-A)Z^{k-1} + A(Z^{k-1} - A^{k-1}).
      \label{eq:93}
    \end{equation}       
Again, from \eqref{eq:93},
\begin{equation*}
    Z^{k-1} - A^{k-1} = (Z-A) Z^{k-2} + A(Z^{k-2}-A^{k-2}).
    \label{eq:94}
\end{equation*}
Thus,
\begin{equation*}
    Z^k - A^k = (Z-A) Z^{k-1} + A(Z-A)Z^{k-2} + A^2(Z^{k-2} - A^{k-2}).
    \label{eq:95}
\end{equation*}
Iteratively,
\begin{multline*}
     Z^k - A^k = (Z-A) Z^{k-1} + A(Z-A)Z^{k-2} +\\ A^2(Z-A)Z^{k-3 }+ \cdots + A^{k-1}(Z-A).
\end{multline*}   
Thus, the $i^{th}$ term will be $A^{i-1}(Z-A) Z^{k-i}$. For $j = i-1$, it becomes $A^j (Z-A) Z^{k-1-j}$. Putting $Z = A+H$, it can be written as $A^j H(A+H)^{k-1-j} $. Hence,
\begin{equation}
   (A+H)^k - A^k = \sum_{j=0}^{k-1} A^j H(A+H)^{k-1-j}, 
   \label{eq:97}
\end{equation}
or, 
\begin{equation*}
    e^{(A+H)}- e^A = \sum_{k=1}^\infty\sum_{j=0}^{k-1} \frac{1}{k!} A^j H(A+H)^{k-1-j}.
    \label{eq:98}
\end{equation*}
For each term, we can write $(A+H)^{k-1-j}$ as $A^{k-1-j} +\{ (A+H)^{k-1-j}- A^{k-1-j}\} $,
\begin{multline}
e^{A+H} - e^A = 
\sum_{k=1}^\infty \sum_{j=0}^{k-1} \frac{1}{k!} A^j H  A^{k-1-j}  \\+ \sum_{k=1}^\infty \sum_{j=0}^{k-1} \frac{1}{k!} A^j H  A^{k-1-j} \{(A+H)^{k-1-j}-A^{k-1-j} \}
    \label{eq:99}
\end{multline}
We identify the first term in the RHS of \eqref{eq:99} as the beta-integral form (Gateaux derivative), and let us denote the remainder term as $R(H)$.
\begin{equation*}
    e^{(A+H)}- e^A - \partial \exp(A;H) = R(H)
    \label{eq:100}
\end{equation*}
Where, 
\begin{equation}
    R(H) = \sum_{k=1}^\infty\sum_{j=0}^{k-1}\frac{1}{k!}A^jH\{ (A+H)^{k-1-j}- A^{k-1-j}\}.
    \label{eq:101}
\end{equation}
From \eqref{eq:97}, we can rewrite \eqref{eq:101} as:
\begin{equation*}
    R(H) = \sum_{k=2}^\infty\sum_{j=0}^{k-2}\sum_{\ell=0}^{k-2-j}\frac{1}{k!} A^j H A^\ell HA^{k-2-j-\ell}.
    \label{eq:102}
\end{equation*}
Let, $m = k-2$, or $k = m+2$, 
\begin{equation*}
    R(H) = \sum_{m=0}^\infty\sum_{j=0}^{m}\sum_{\ell=0}^{m-j}\frac{1}{(m+2)!}A^jHA^\ell HA^{m-j-\ell}.
    \label{eq:103}
\end{equation*}
For any submultiplicative matrix norm, we have,
\begin{multline*}
    ||A^jHA^\ell HA^{m-j-\ell}|| \leq 
    ||A^j||\cdot||H|| \cdot||A^\ell||\cdot||H||\cdot \\||A^{m-j-\ell}||
    =||A||^m\cdot||H||^2.
    \label{eq:104}
\end{multline*}
In the sums over $j$ and $\ell$, we see that for $j = 0, 1, 2, \dots m$, we have $\ell= 0, 1, 2, \dots (m-j)$. Thus, total number of terms will be,
\begin{equation*}
    \text{Number of terms} = \sum_{j=0}^{m}(m-j+1) = \frac{(m+1)(m+2)}{2}.
    \label{eq:105}
\end{equation*}
Now, we have,
\begin{equation*}
    ||R(H)|| \leq \sum_{m=0}^\infty \frac{(m+1)(m+2)}{2(m+2)!} ||A||^m\cdot||H||^2,
    \label{eq:106}
\end{equation*}
or, 
\begin{equation*}
    ||R(H)|| \leq \frac{1}{2}||H||^2\sum_{m=0}^\infty \frac{1}{m!} ||A||^m,
    \label{eq:106}
\end{equation*}
or,
\begin{equation*}
    ||R(H)|| \leq \frac{1}{2}||H||^2e^{||A||},
    \label{eq:107}
\end{equation*}
or,
\begin{equation*}
    \frac{||R(H)||}{||H||} \leq \frac{1}{2}||H|| e^{||A||}.
    \label{eq:108}
\end{equation*}
As $||H|| \to 0$, $\frac{||R(H)||}{||H||}\to 0$, and,
\begin{equation*}
    R(H) = o(||H||).
    \label{eq:109}
\end{equation*}
Thus, the Fréchet derivative exists, and from \eqref{eq:90},
\begin{equation*}
    D\exp(A;H) = \int_0^1 e^{(1-s)A}H e^{sA} ds.
    \label{eq:110}
\end{equation*}

\section{Concluding Remarks}
\label{conclusion}
In this report, we derived the time derivative of the matrix exponential $\exp(A(t))$ using Taylor expansions, Euler integrals and Lie algebraic tools. We demonstrated the equivalence between the integral, Lie bracket, and adjoint operator forms. The differential equation approach of Rossmann (and Wilcox) has the limitation that it depends on knowing the form of the answer. Our approach using the beta integral is deductive and free from such functional form assumptions, as shown in Section \ref{lie}. The results give explicit formulas for the Fréchet differential of the
exponential map.

\footnotesize
\section{Acknowledgements}
Professor P. S. Krishnaprasad suggested the problem stated in Section \ref{intro} \--- to compute the derivative of
exponential of a time-parametrized curve in a matrix space. After I derived the form (4), he suggested
that I work out an integral form for this expression. This I did using the beta integral, leading to the form
(9). He then suggested that I consider the derivation in Rossmann’s book, which naturally led me to the
calculations that followed (9). Moreover, he reviewed the first draft and suggested helpful corrections.

Matthew S. Hankins helped me rectify a mistake I initially made while showing the existence of the Fréchet differential in Section \ref{frechet}. His insights regarding the \LaTeX{} formatting were valuable as well.

\end{document}